\documentclass[aps,prc,showpacs,amssymb,amsmath,amsfonts,nofootinbib,floatfix]{revtex4-2}
\usepackage{graphicx}

\makeatletter
\def\p@subsection{}

\def\p@subsubsection{}
\makeatother
\usepackage{natbib}
\usepackage[usenames]{color}
\usepackage{epsfig,graphicx}
\usepackage{epstopdf}
\usepackage{amssymb,amsmath,eqparbox}
\usepackage{subfig}
\usepackage{epstopdf}
\usepackage{enumitem}
\usepackage{comment}
\usepackage[bookmarks={false}]{hyperref}
\usepackage{caption}
\def \red{\color{red}}

\definecolor{grey}{rgb}{0.9,0.9,0.9}

\definecolor{black}{rgb}{0,0,0}

\hypersetup{pdfstartview={XYZ null null 1.}}

\def \AA/PWA{Svarc2020,Svarc2022}

\newcommand{\be}{\begin{eqnarray}}
\newcommand{\ee}{\end{eqnarray}}
\newcommand{\bc}{\begin{center}}
\newcommand{\ec}{\end{center}}

\def \AA/PWA{Svarc2020,Svarc2022}

\newcommand{\RudjerBoskovic}{Rudjer Bo\v{s}kovi\'{c} Institute, Bijeni\v{c}ka cesta 54, P.O. Box 180, 10002 Zagreb, Croatia}
\newcommand{\Tesla}{Tesla Biotech d.o.o., Mandlova 7, 10000 Zagreb, Croatia}

\begin{document}

\allowdisplaybreaks

\title{Single-channel, single-energy partial-wave
analysis with continuity improved
through minimal phase constraints}
\author{A.~{\v{S}}varc}\email[Corresponding author: ]{svarc@irb.hr}
\affiliation{\RudjerBoskovic,\Tesla}

\author{R.L. Workman}
\affiliation{Institute for Nuclear Studies, Department of Physics, \
The George Washington University, Washington D.C. 20052, USA}

\date{\today}

\begin{abstract}
\vspace*{1.cm}

Single-energy partial-wave analysis has often been applied as a way to fit data with minimal model dependence. However, remaining unconstrained, partial waves at neighboring energies will vary discontinuously because the overall amplitude phase cannot be determined through single-channel measurements. This problem can be mitigated through the use of a constraining penalty function based on an associated energy-dependent fit. However, the weight given to this constraint results in a biased fit to the data. In this paper, for the first time, we explore a constraining function which does not influence the fit to data. The constraint comes from the overall phase found in multi-channel fits which, in the present study, are the Bonn-Gatchina and J\"ulich-Bonn multi-channel analyses. The data are well reproduced and weighting of the penalty function does not influence the result. The method is applied to $K \Lambda$  photoproduction data and all observables can be maximally well reproduced. While the employed multi-channel analyses display very different multipole amplitudes, we show that the major difference between two sets of multipoles can be related to the different overall phases.

\end{abstract}

\maketitle

\newpage
\section{Introduction}

Meson-nucleon scattering and meson photoproduction have been extensively studied over the last decades in a comprehensive joint program of experiments and theoretical studies principally at ELSA, GRAAL, JLab and MAMI. Multi-channel theory approaches, attempting a simultaneous fit to the world datasets of dominant open channels, have been carried out and updated by the Bonn-Gatchina~\cite{BoGa}, J\"ulich-Bonn~\cite{JuBo}, Kent State University~\cite{Kent}, and MTZ~\cite{MTZ} groups among others. Some single-channel analyses have also remained active,  in particular GWU-SAID~\cite{GWU-SAID} and MAID~\cite{MAID}.
No longer active but historically important analyses, focusing on pion-induced reactions, were carried out by the Karlsruhe-Helsinki group~\cite{Hoehler}  as a single-channel analysis,
and Carnegie-Mellon-Berkeley group~\cite{CMB} as a coupled-channel one.
Coupled-channel theoretical models are very powerful as they employ the constraint of multi-channel unitarity, but may be very computationally intensive and can have very different starting assumptions. This has made comparisons more difficult.
Single-channel energy-independent approaches attempt to reduce theoretical input to a minimum, and aim to extract partial waves or multipoles directly from the data, using as few assumptions as possible. However, they all face the well known problem that single-channel single-energy partial-wave analysis, in the inelastic region, gives discontinuous results due to the invariance of single-channel observables with respect to an overall phase change (continuum ambiguity, see ref.~\cite{Continuum-ambiguity}). Thus, some constraints have to be introduced, and this introduces model-dependence into the analysis. In this paper we analyze the constraints used to achieve the needed continuity, and propose to use the minimal constraint which does not change single-channel observables. This also leads to a way of comparing different multi-channel results.
\\ \\ \indent
Before proceeding, we first briefly rephrase the existing problem with single-channel single-energy analyses due to the overall phase ambiguity.
Since the observables can be written in terms of bilinears (such as the real or imaginary part of $b_i b_j^*$, with $i,j$ running from 1 to 4 for pseudo-scalar photoproduction),
resulting in real valued functions, the multiplication of all amplitudes by
a common phase has no effect on observables. Hence, only absolute values and all three relative phase angles of reaction amplitudes are uniquely determined; the overall phase always remains unknown.
This has a profound effect on our results: any of our solutions, whether they be reaction amplitudes or partial waves, are  manifestly non-unique.
In fact, we have an infinite number of equivalent solutions with different overall phase (if our solution is considered to be a set of four reaction amplitudes with four absolute values and four phase angles). For more details see Tables~\ref{tab:PhotoproductionObservables} and \ref{tab:Relamp1}. To proceed, we need some way of fixing the unmeasurable overall phase, and that is done by introducing constraints in single-energy partial wave analyses.
\\ \\  \indent
 Very often the employed constraints come from some theoretical model in the form of theoretical reaction amplitudes. However, in that case, single-energy  analyses become strongly model dependent, as constraining functions directly influence the fit to observables. One of the first constraining methods was used by the Karlsruhe - Helsinki partial wave analysis~\cite{Hoehler} in the form of fixed-t analyticity. The method consisted in introducing a penalty function as a constraint which required that, simultaneously with fitting the data, reaction amplitudes had to be very close to those obtained by a Pietarinen expansion in the t-variable of a single-channel, and this constraint imposed analyticity of amplitudes in the s-variable. In that way, continuity in phase was automatically achieved, but a much stronger constraint to the form of constraining amplitudes was also imposed. Obviously, this constraining method influenced the single-channel observables. A very similar way of constraining single-channel, single energy partial wave analysis has been used by Zagreb group~\cite{Svarc2020,Svarc2022} where analyticity in t-variable has been replaced by analyticity in s-variable. The method was successful but the main problem with the constraining function influencing observables remained.
\\ \\ \indent
Obviously, the main problem of these approaches was that as constraining functions they were using full amplitudes which influence the observables.  The first attempt to use only a phase\footnote{which has no influence onto observables} to achieve continuity of single-channel, single-energy partial wave analysis was introduced by Grushin~\cite{Grushin89} and  used by Bonn group~\cite{Phase}. The method consisted in fixing the phase of one of the multipoles to zero, usually the $E_{0+}$ multipole.  In practice this was achieved by imposing $Im(E_{0+})=0$, $Re(E_{0+})>0$. Unfortunately, this method yields a very restrictive form of the chosen multipole. So, until recently, there existed no general way to constrain a single-channel, single-energy partial wave analysis without either influencing the fit to observables or fixing the values of some multipoles.
\\ \\ \indent
A starting point for the present method was discussed in Ref.~\cite{Svarc2018} for the complete set of observables created as numeric data with infinite precision in $\eta$ photoproduction. There it was shown how multipoles explicitly depend on the overall phase, and that their continuity can be achieved if the overall phase is fixed to some predetermined value with no influence on any single-channel observable.
\\ \\ \indent
In this paper, we apply this finding to experimental data (with uncertainties), in the data base for $K \Lambda$ photoproduction. Our constraining function was chosen to be the overall transversity amplitude phase, and the constraining procedure was chosen to be the penalty function technique.  In the Appendix, it is shown that fixing the overall phase is equivalent to fixing the phase of one of the reaction amplitudes, so as a constraining function we take the transversity amplitude $b_1$ phases from two coupled-channel models, Bonn-Gatchina~\cite{BoGa} and Juelich-Bonn~\cite{JuBo}. We show that these phases are significantly different, and thus generate two very different sets of  multipoles. However, by comparing the agreement of calculated observable values with the corresponding experimental ones, for both phase choices, we find that both sets do similarly well in reproducing the world $K \Lambda$ photoproduction data base.

\section{Formalism}
The simplest way to fix the phase and obtain a unique SE PWA is to use the penalty function technique to introduce an additional phase constraint. As it is irrelevant which amplitude phase we fix, we have chosen the phase of amplitude $b_1$. We name it $\Phi_1$, and we fix it to the chosen value $\Phi_1^{pen}$.
\be
\label{Eq1}
\chi^2(W) & = & \sum_{i=1}^{N_{data}}w^i \left[ {\cal O}^{exp}_i (W,\Theta_i) - {\cal O}^{fit}_i \left[{\cal M}^{fit}(W),\Theta_i\right] \right]^2
+ {\cal P}_{1}  \\ \nonumber
 {\cal P}_{1}   &=& \lambda_{ph} \sum_{i=1}^{N_{data}} \left|\Phi_1^{fit}(W,\Theta_i)- \Phi_1^{pen}(W,\Theta_i) \right|
\ee
where ${\cal O}(W,\Theta)$ and  ${\cal M}(W)$ are the generic names for all observables and  multipoles, and $w^i$ is a statistical weight.
In practice, we avoid the difficulties associated with fitting a multi-valued phase, instead using normalized amplitudes in defining
the penalty function
\be
\label{Eq2}
\chi^2(W) & = & \sum_{i=1}^{N_{data}}w^i \left[ {\cal O}^{exp}_i (W,\Theta_i) - {\cal O}^{fit}_i \left[ {\cal M}^{fit}(W),\Theta_i \right] \right]^2
+ {\cal P}_{1}  \\ \nonumber
 {\cal P}_{1}   &=& \lambda_{ph} \sum_{i=1}^{N_{data}} \left| \frac{b_{1}^{fit}(W,\Theta_i)}{|b_{1}^{fit}(W,\Theta_i)|}- \frac{b_{1}^{pen}(W,\Theta_i)}{|b_{1}^{pen}(W,\Theta_i)|} \right|.
\ee
This replaces the fit to a phase with a fit to its sine and cosine, a better-behaved penalty function.

\section{Illustration of single-channel PWA applied to $K \Lambda$ photoproduction data}
Single-energy analysis has often been used to search for systematic deviations from an underlying energy-dependent fit, searching for missing structure. However, this method can bias the single-energy results as a large penalty function can generate single-energy values arbitrarily close the the energy-dependent input. The idea to use theoretical constraining amplitudes has already been tried in past~\cite{Hoehler,Svarc2020,Svarc2022}, but using using the T-matrix  as a constraining function influenced the fit to these observables.   Below, we describe a fit to data constrained only by the overall phase taken from a multi-channel analysis, which cannot be directly measured and so has no effect on the quality of fit to data.

\subsection{The $\gamma p \to K^+ \Lambda$ Data Base} \label{Data:base}
The $\gamma p\to K^+ \Lambda$ data base, used in this study, is identical to one fitted in Ref.~\cite{Svarc2022}.
In Table~\ref{tab:expdata} our data base is summarized. It has been taken, in numerical form, from the Bonn-Gatchina and George-Washington-University (SAID) web pages~\cite{BG-web,GWU-web} and interpolated to produce a grid of common energy/angle points:
\begin{table*}[htb]
\begin{center}

\caption{\label{tab:expdata} Experimental data from CLAS, and GRAAL used in our PWA. Note that the observables $C_x$ and $C_z$
are measured in a rotated coordinate frame~\cite{Bradford}. They are related to the standard observables $C_{x'}$ and $C_{z'}$ in
the $c.m.$ frame by an angular rotation: $C_x= C_{z'} \sin(\theta)+ C_{x'} \cos(\theta)$, and $C_z= C_{z'} \cos(\theta)- C_{x'}
\sin(\theta)$, see Ref.~\cite{Anisovich2017a}. }
\bigskip
\begin{ruledtabular}
\begin{tabular}{ccccccc}
 Obs.        & $N$ & $E_{c.m.}$~[MeV] & $N_E$  & $\theta_{cm}$~[deg] & $N_\theta$ & Reference    \\
\hline
 $d\sigma/d\Omega \equiv \sigma_0$ & $3615$ & $1625-2295$ & $268$  & $28 - 152$ & $5-19$ & CLAS(2007)~\cite{Bradford}, CLAS(2010)~\cite{McCracken} \\
 $\Sigma$   & $ 400$ & $1649 - 2179$ & $ 34$  & $35 - 143$ & $6-16$ & GRAAL(2007)~\cite{Lleres}, CLAS(2016)~\cite{Paterson} \\
 $T$        & $ 408$ & $1645 - 2179$ & $ 34$  & $31 - 142$ & $6-16$ &  GRAAL(2007)~\cite{Lleres},CLAS(2016)~\cite{Paterson}  \\
 $P$        & $ 1597$ & $1625 - 2295$ & $ 78$  & $28 - 143$ & $6-18$ & CLAS(2010)~\cite{McCracken},  GRAAL(2007)~\cite{Lleres} \\
 $O_{x'}$   & $ 415$   & $1645 - 2179$ & $ 34 $  & $31 - 143$ & $6-16$ & GRAAL(2007)~\cite{Lleres}, CLAS(2016)~\cite{Paterson} \\
 $O_{z'}$   & $ 415$  & $1645 - 2179 $ & $ 34 $  & $31 - 143$ & $6-16$ &  GRAAL(2007)~\cite{Lleres}, CLAS(2016)~\cite{Paterson} \\
 $C_x$     & $ 138$  & $1678 - 2296$ & $ 14 $  & $31 - 139$ & $9 $ &  CLAS(2007)~\cite{Bradford} \\
 $C_z$      & $ 138 $ & $1678 - 2296 $ & $ 14 $  & $31 - 139$ & $9$ &  CLAS(2007)~\cite{Bradford}
\end{tabular}
\end{ruledtabular}
\end{center}
\end{table*}
for general details related to the 2-dimensional interpolation and its implementation, see Refs.~\cite{Svarc2020,Svarc2022}. However, the interpolating/extrapolating stability in the present study is significantly improved with respect to Refs.~\cite{Svarc2020,Svarc2022}. Observe that, in angular range, not all measured observables overlap, and for some data groups extrapolations are needed. However, this extrapolation at extreme forward and backward angles can become rather ambiguous if it is completely determined by the fitting software. Therefore, we have introduced additional kinematical constraints to the measured data at the angular limits:
\be
\Sigma = P = T = O_{x'} = O _{z'}  =  C_{x} = 0 \hspace{0.7cm}  \& \hspace{0.7cm} C_{z} = 1 \hspace{0.7cm} {\rm at} \hspace{0.7cm}  \cos \theta  = \pm 1
\ee
For the differential cross section $d\sigma/d\Omega$, the Bonn-Gatchina theoretical values were used as a constraint at these extreme angles.
This stabilizes the extrapolations at forward and backward angles significantly, and enables us to increase the angular range from experimentally measured -0.7 $< \cos \theta <$ 0.8 to a broader -0.9 $ < \cos \theta <$ 0.9, improving the reliability of a partial wave reconstruction.

However, note that in spite of the fact that we have eight measured observables, this set is still not complete (see Ref.~\cite{Chiang:1996em}).
Namely, observables $O_{x'}, O_{z'}, C_x, {\rm and} \, C_z$ are determining the same two relative angles $\phi_{14}$ and $\phi_{23}$ (see Table~\ref{tab:PhotoproductionObservables}),
while the third relative angle remains undefined. This set, when fitted, produces results with large scatter, leading us to consider use of an expanded set of observables.
 To obtain a complete set of observables, we have added two  $\cal{B} \cal{T}$ observables (see again  Ref.~\cite{Chiang:1996em}).
 As they are not measured, we have taken them as pseudo-data from a theoretical model. We acknowledge that this introduces additional theory dependence, but at this exploratory stage it appears necessary.
 Thus, we introduce pseudo-data E and F generated either by the Bonn-Gatchina or the J\"ulich-Bonn model. We first generate numeric data from the respective models, then
 randomize them by 10~\%, and attribute to them a 10~\% error. In this way we obtain quasi-realistic data set for two additional observables.
 The obtained set of ten observables is beyond that required for a complete experiment.
 We could drop a pair a pair of connected observables (either $O_{x'} \, {\rm and} \, O_{z'}$  or $ C_x \, {\rm and} \, \, C_z$, but we have opted to fit the over-defined set of 10 observables,
 to the maximize the link to experiment. Further tests in the direction of self-consistency of the observables $O_{x'}, O_{z'}, C_x, {\rm and} \, C_z$ can be done.

\subsection{\textbf{\emph{Exploring the overall phase}}}
\begin{figure}[h!]
\bc
\includegraphics[width=0.45\textwidth]{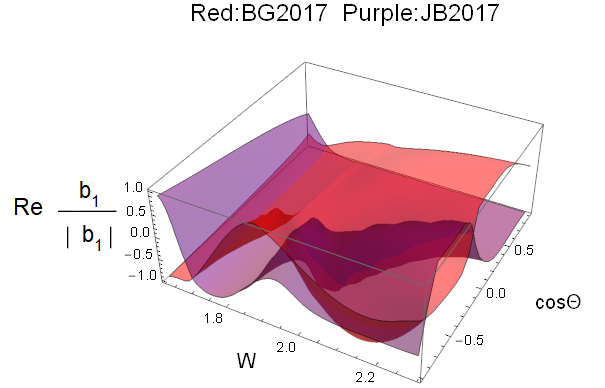} \includegraphics[width=0.45\textwidth]{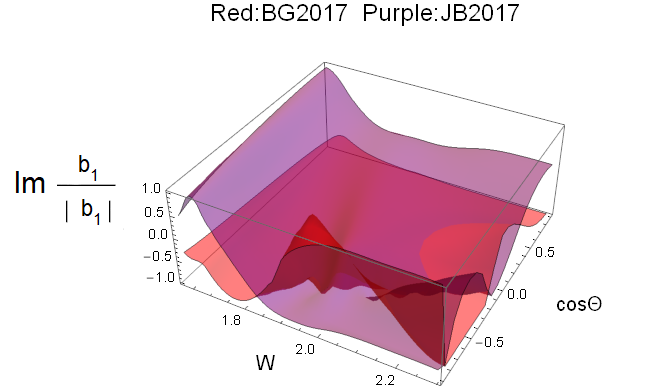} \\

\caption{\label{Comparison-of-phases}(Color online) The comparison of BG2017 and JuBo2022 $b_1$  phases.}
\ec
\end{figure}

As stressed in the Introduction, single-channel, single-energy  analyses cannot be made theory independent, as the overall phase in
inelastic channels cannot be measured from data. As described in papers on the continuum ambiguity, a way to determine this phase is
to use multi-channel unitarity.
\be
Im \, T_{ab} &=& \sum_{c=1}^{all \, channels} {T_{ac}^* \, \rho_c \, T_{cb}}
\label{MCunit}
\ee
where $a,b,c$ are channel indices, $T_{ab}$ is the T-matrix of the reaction $a \rightarrow b $ and $\rho_c$ is the phase space for the channel $c$. This equation is just the mathematical expression of conservation of probability for multi-channel reactions. This equation restores the connection between overall phases of analyzed channels.
\\ \\ \indent
It is clear that multi-channel unitarity constraints can only be obtained from coupled-channel fits which have enforced this in the model used.
We have chosen to fix the overall phase of the $b_1$ transversity amplitude to the values obtained in the Bonn-Gatchina and J\"ulich-Bonn models, and have compared results.
A comparison of phases is given in Fig.~\ref{Comparison-of-phases}.
Clearly the phases coming from the Bonn-Gatchina and J\"ulich-Bonn models do not agree, so we should expect that multipoles will show sizable differences.
This is confirmed as we compare results from the two multi-channel fits in the next section.
\\ \\
\clearpage
 \subsection{Fitting an over-complete set of observables with BG2017 phase and BG2017 pseudo-data} \label{BG2017fit}
 Below are results for dominant multipoles obtained by fitting the full set of data, consisting of 8 experimentally measured observables and 2 pseudo-observables, over the energy range $1625 \le W \le 2179.83$ MeV. The agreement between
 energy-dependent BG2017 multipoles and our single-energy values is very good in most waves.
 \begin{figure}[h]

\bc
\includegraphics[width=0.32\textwidth]{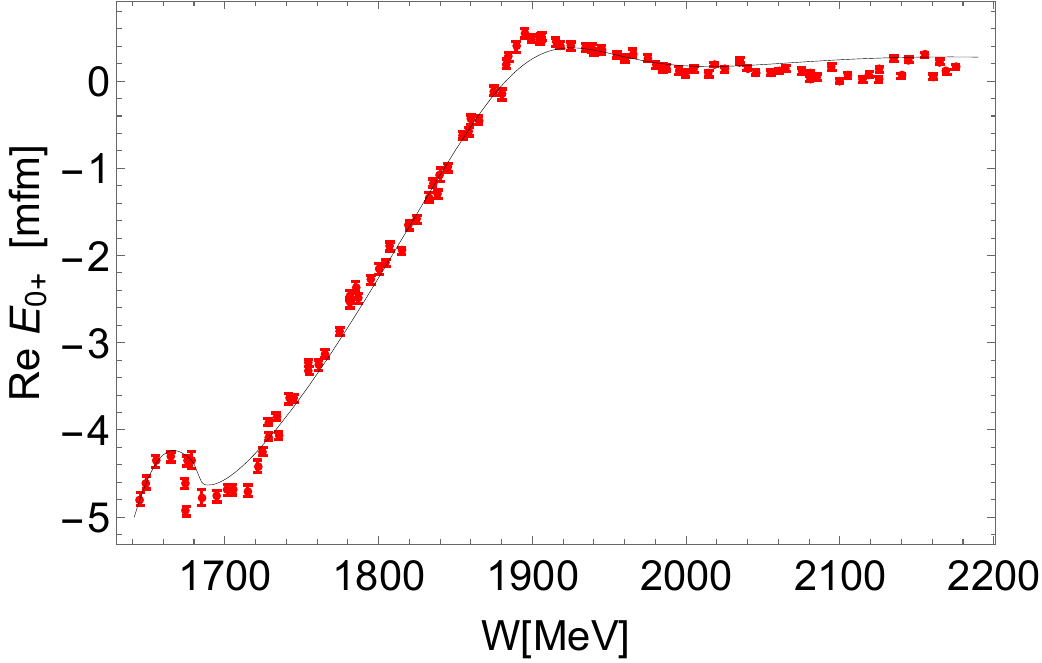} \hspace{0.5cm}
\includegraphics[width=0.32\textwidth]{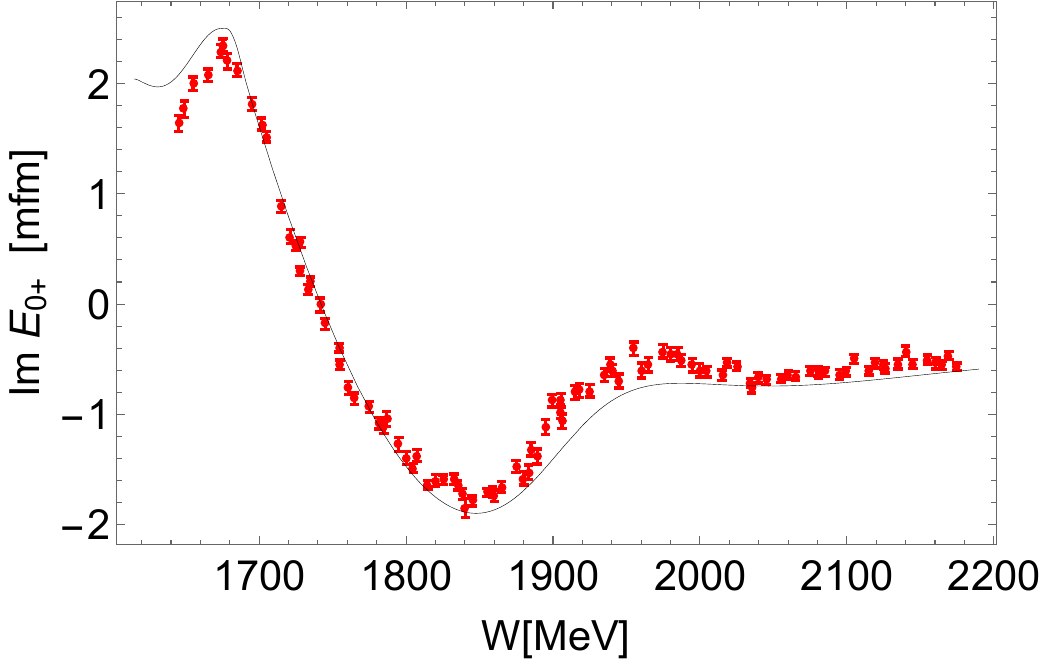}  \\
\includegraphics[width=0.32\textwidth]{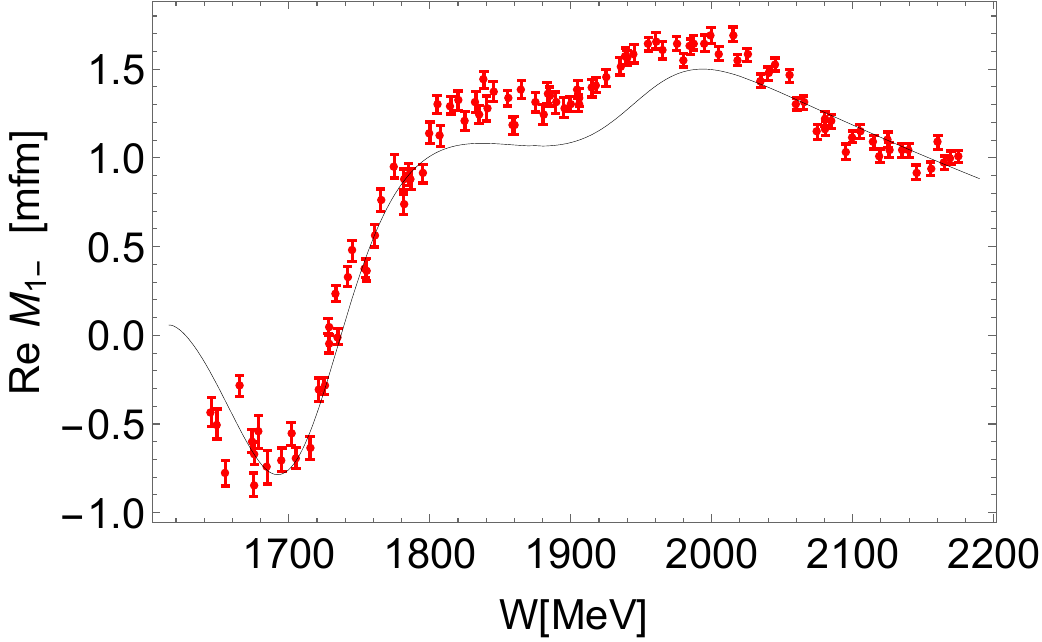} \hspace{0.5cm}
\includegraphics[width=0.32\textwidth]{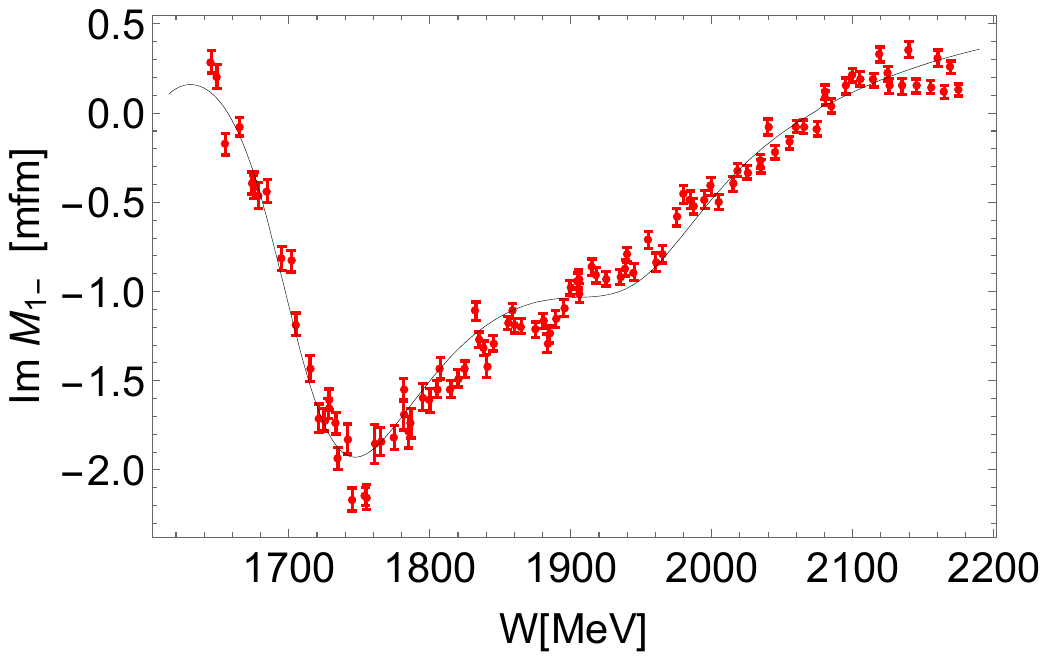}  \\
\includegraphics[width=0.32\textwidth]{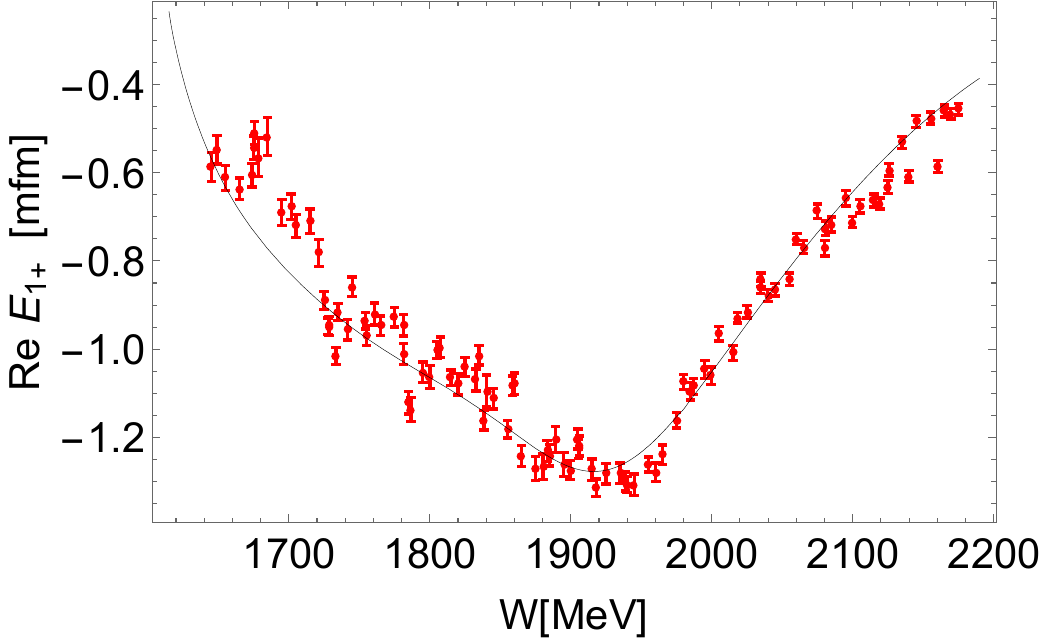} \hspace{0.5cm}
\includegraphics[width=0.32\textwidth]{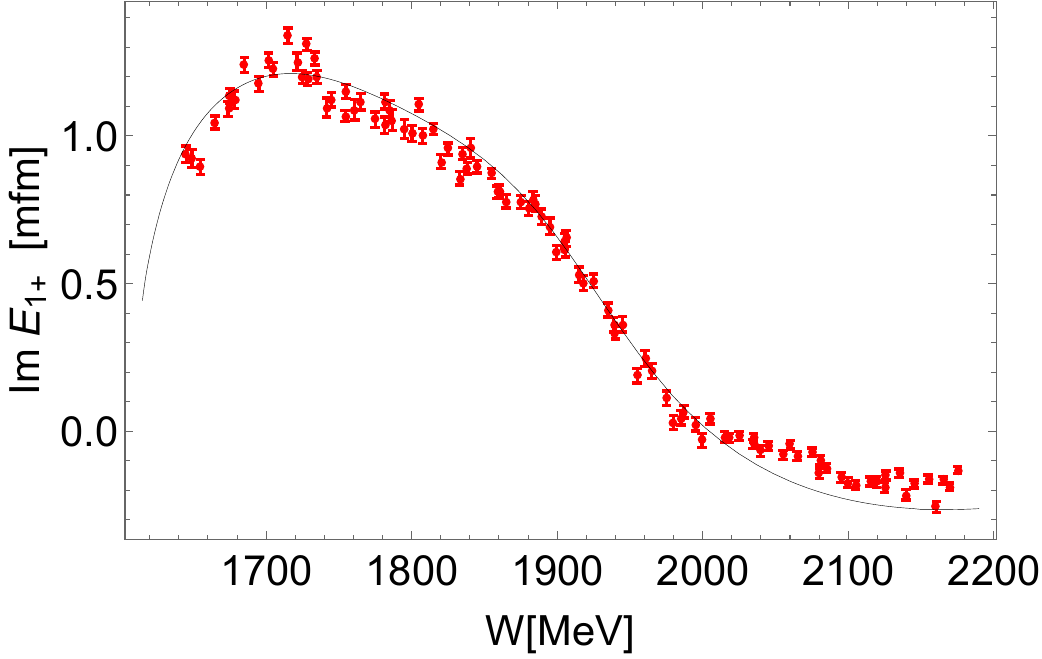}  \\
\includegraphics[width=0.32\textwidth]{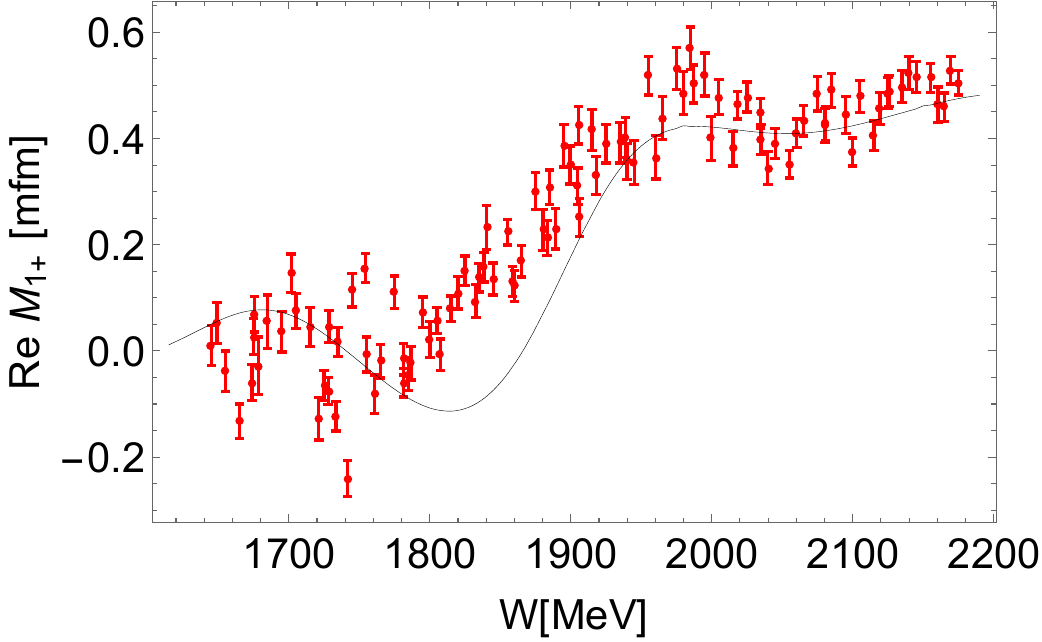} \hspace{0.5cm}
\includegraphics[width=0.32\textwidth]{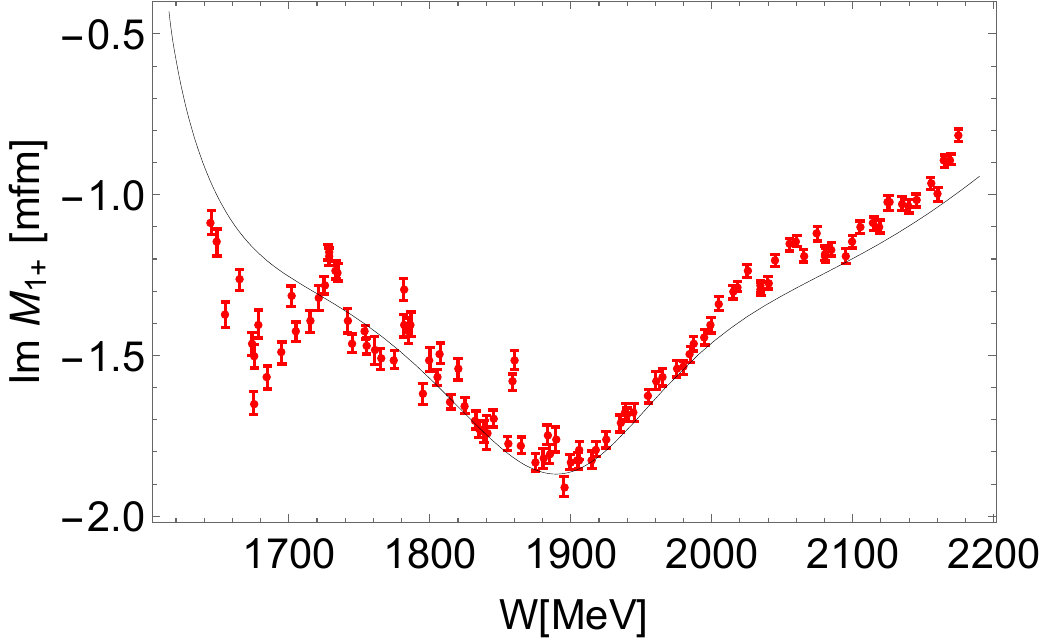}  \\
\includegraphics[width=0.32\textwidth]{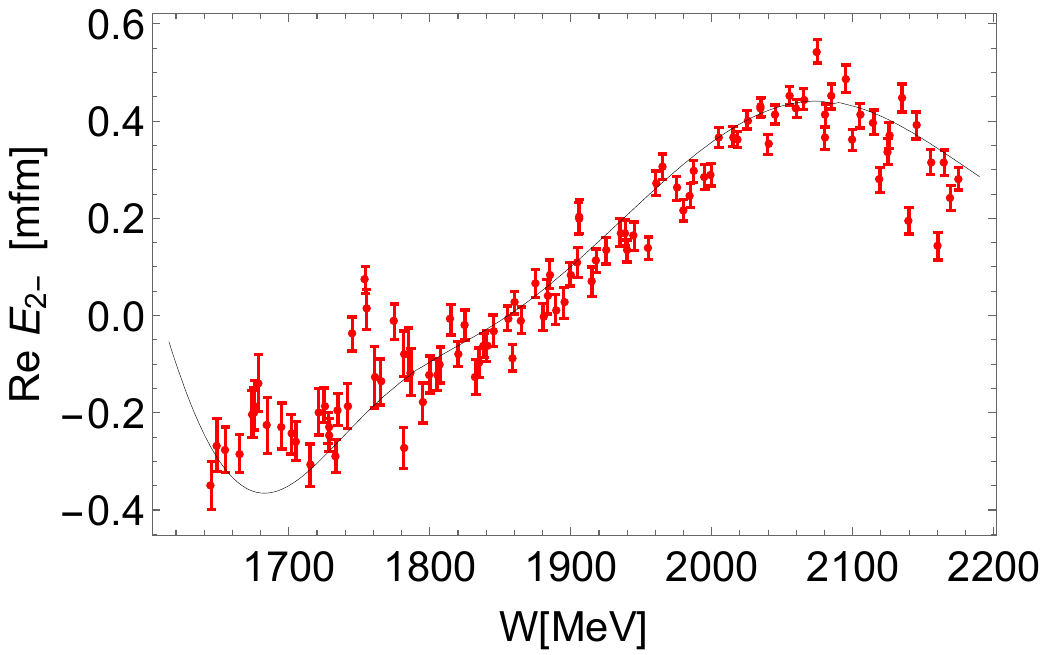} \hspace{0.5cm}
\includegraphics[width=0.32\textwidth]{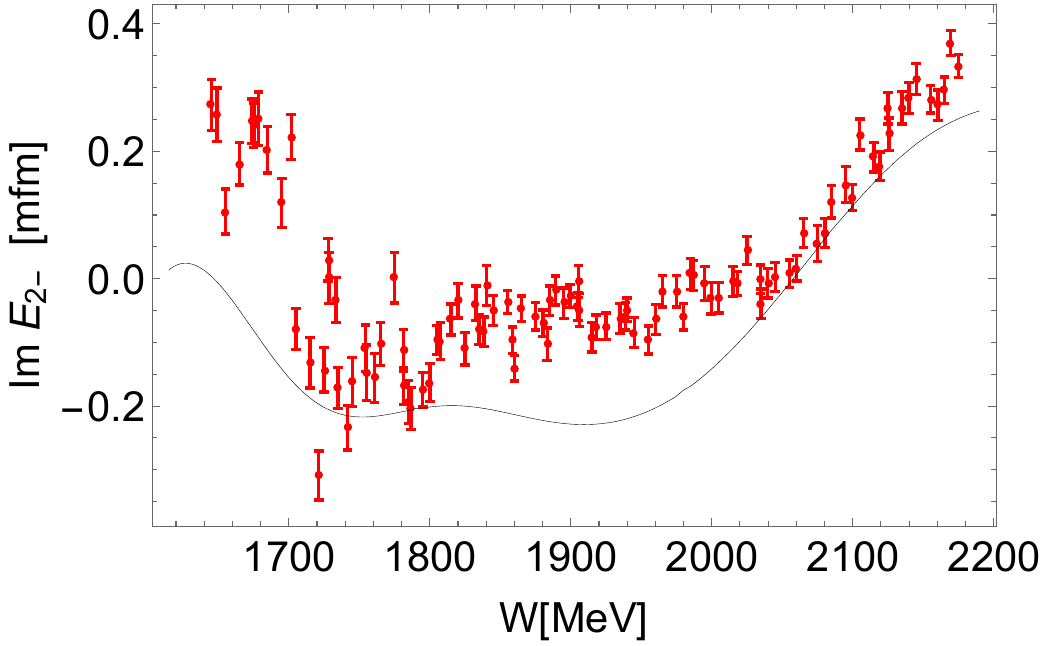}  \\
\caption{\label{MultipolesBG:a}(Color online) The multipoles for the $L=0$, $1$ and $2$ partial waves. Red discrete symbols correspond to the single-channel PWA,  and the full black line gives the BG2017 energy-dependent solution for comparison.  }
\ec
\end{figure}

\clearpage \noindent
Having seen the level of agreement between multipoles, given by BG2017 and by our single-channel single-energy approach, we compare the
two solutions at the level of total $\chi^2$ per degree of freedom.
The black and red lines connect total $\chi^2$ values from the BG2017 and single-energy fits respectively. The $\chi^2$ is a sum of contributions
from the set of observables at each interpolation energy. The single-energy values oscillate around the formally expected value of unity.
The energy-dependent BG2017 values show more scatter and higher values, as should be expected, given that a different (extended and multi-channel) database was used.

\begin{figure}[h!]
\bc
\includegraphics[width=0.6\textwidth]{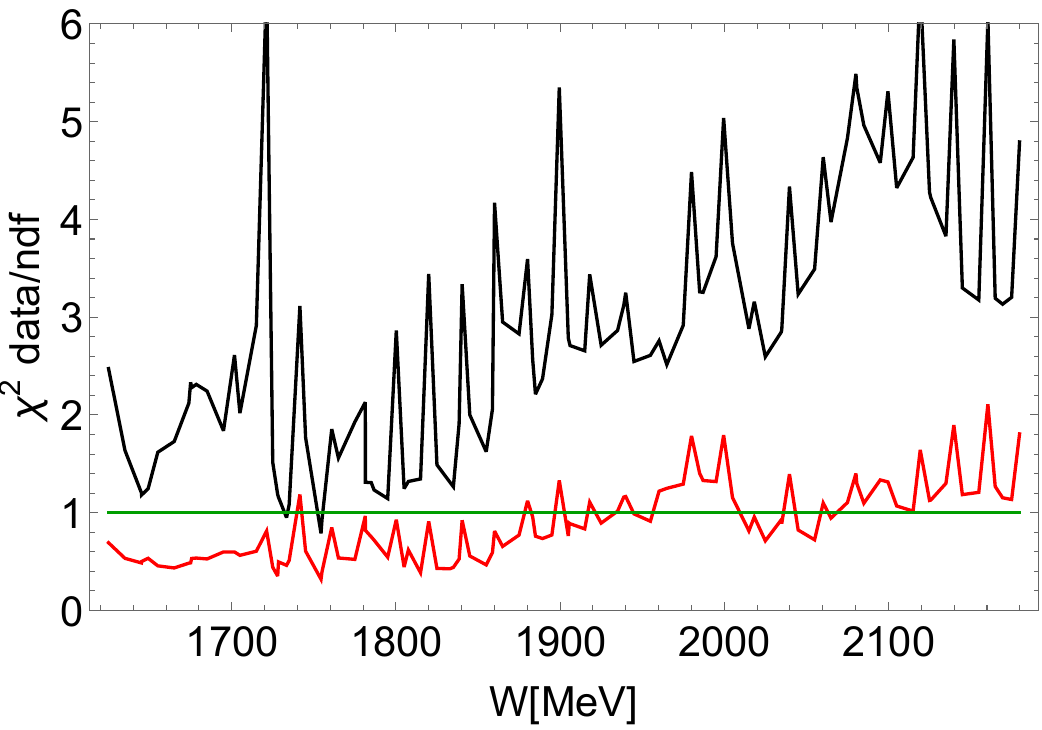}  \\
\caption{\label{chi2BG}(Color online) The obtained total $\chi^2$ per degree of freedom.  Black color denotes BG2017 values, and red color denotes our model.}
\ec
\end{figure}
\clearpage
 \subsection{Fitting an over-complete set of observables with JB2017 phase and JB2017 pseudo-data} \label{JB2017fit}
 Having applied the single-energy method to BG2017, we next check to see if consistent results follow
 with the use of a very different multi-channel technique in constraining the overall phase.
Below are the results for dominant multipoles coming from a fit to the complete set of data consisting of 8 measured observables and 2 pseudo-observables, obtained from the
J\"ulich-Bonn fit (JB2017) in the energy range $1625 \le W \le 2179.83$ MeV. Again, the comparison between energy-dependent and single-energy multipoles is
reasonable, with the closest agreement away from threshold.
\begin{figure}[h!]
\bc
\includegraphics[width=0.32\textwidth]{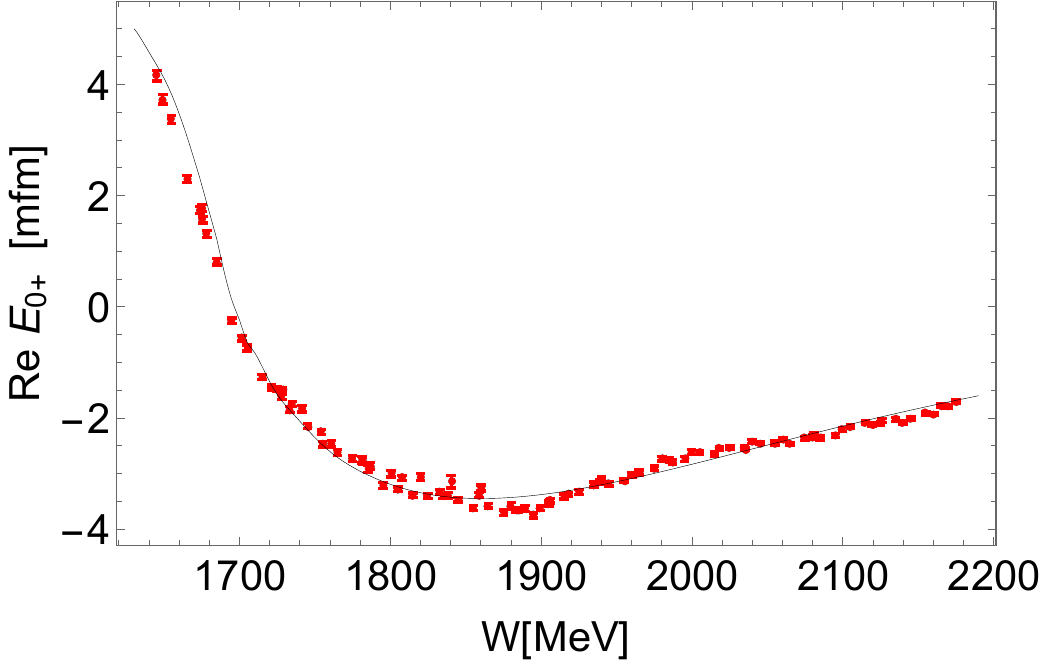} \hspace{0.5cm}
\includegraphics[width=0.32\textwidth]{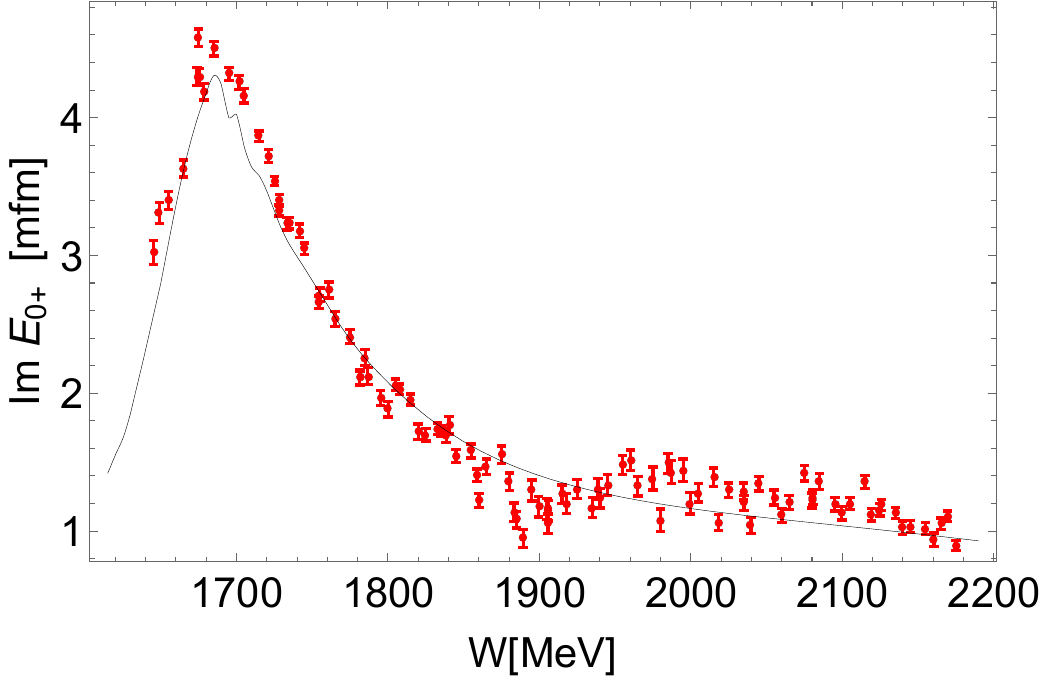}  \\
\includegraphics[width=0.32\textwidth]{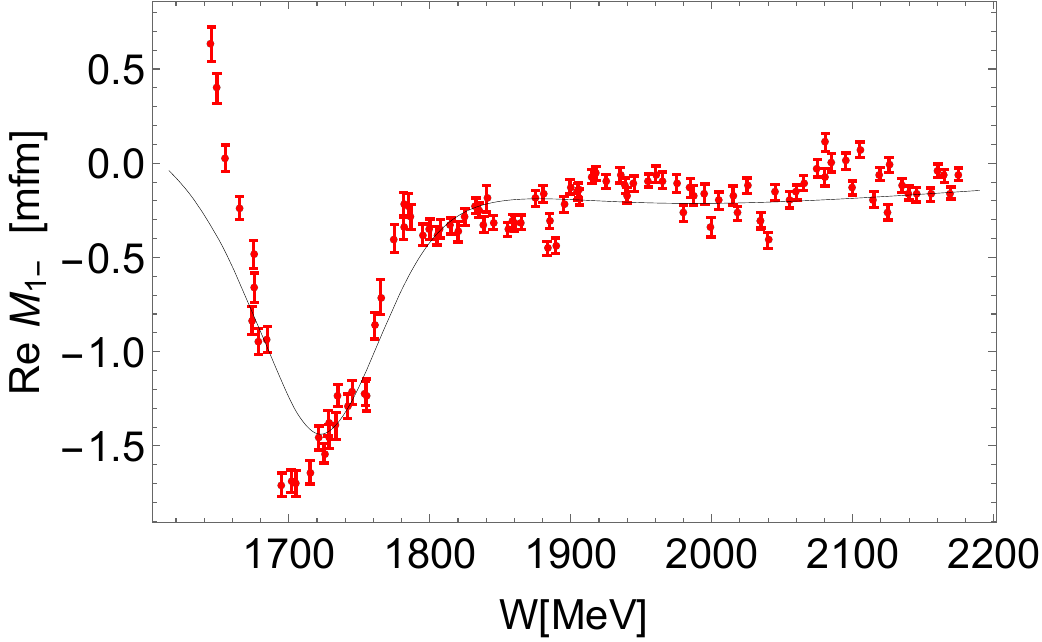} \hspace{0.5cm}
\includegraphics[width=0.32\textwidth]{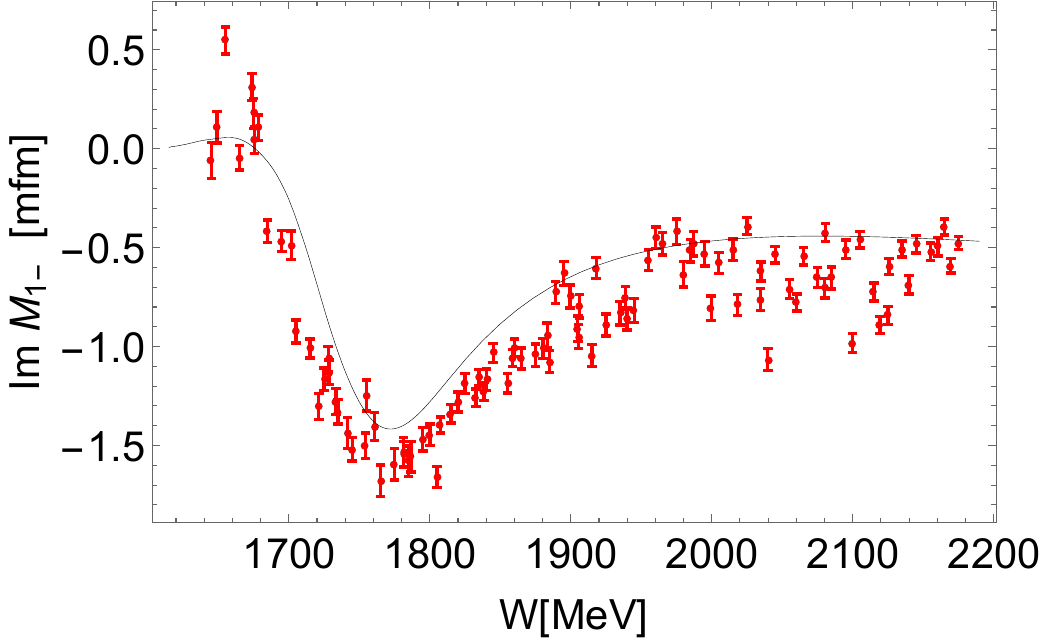}  \\
\includegraphics[width=0.32\textwidth]{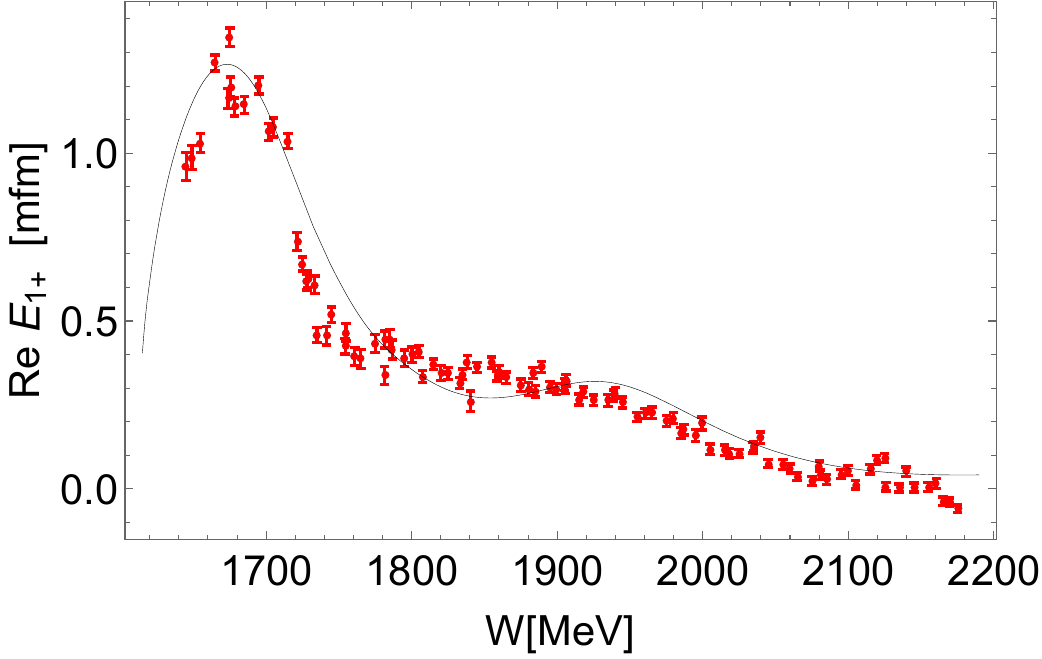} \hspace{0.5cm}
\includegraphics[width=0.32\textwidth]{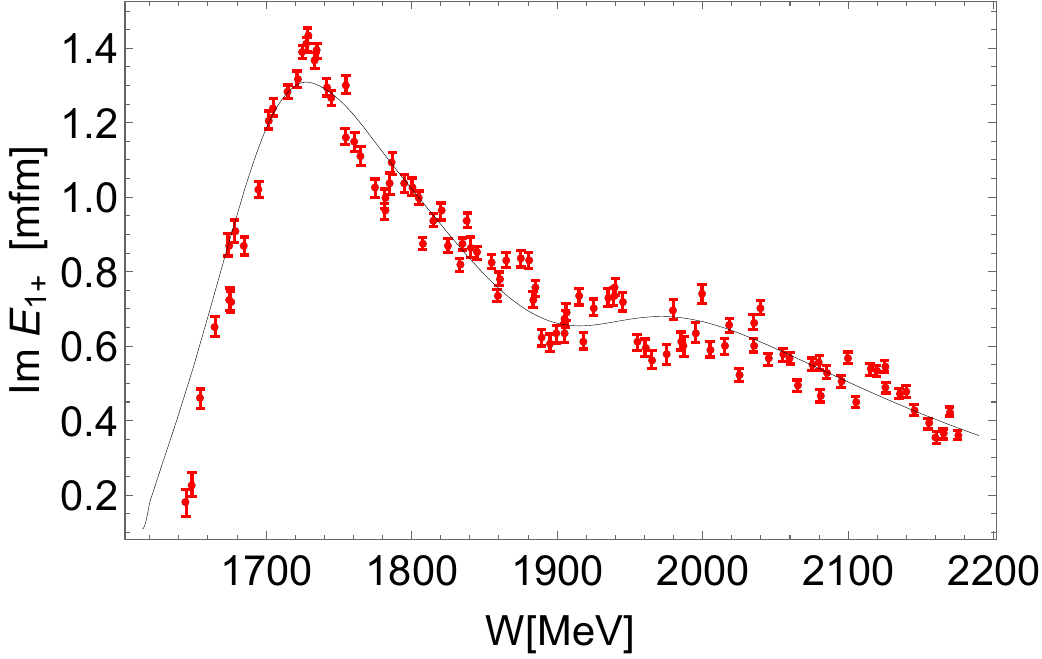}  \\
\includegraphics[width=0.32\textwidth]{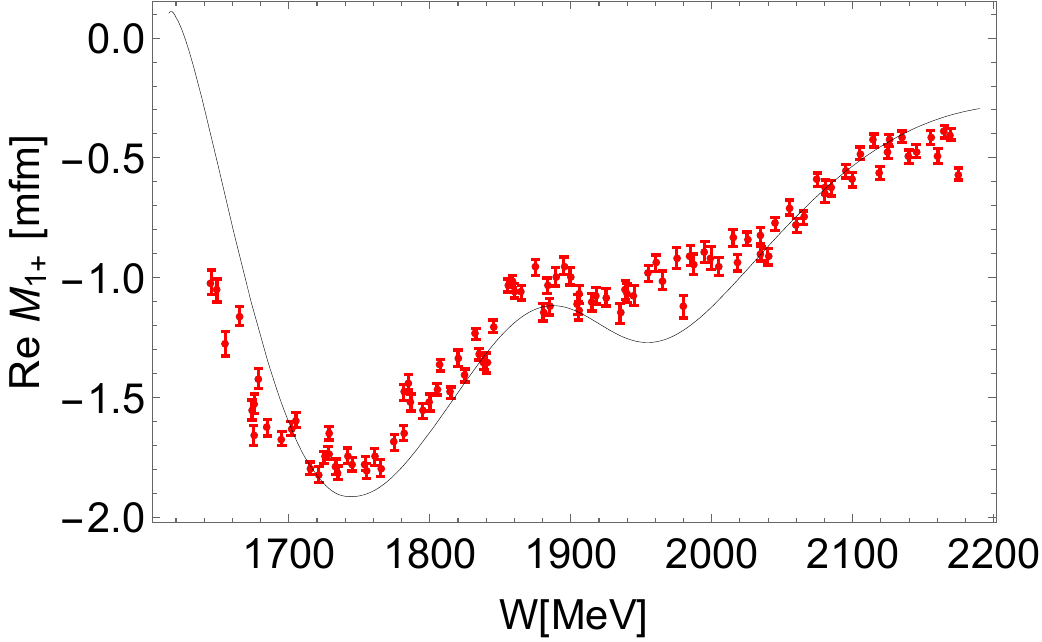} \hspace{0.5cm}
\includegraphics[width=0.32\textwidth]{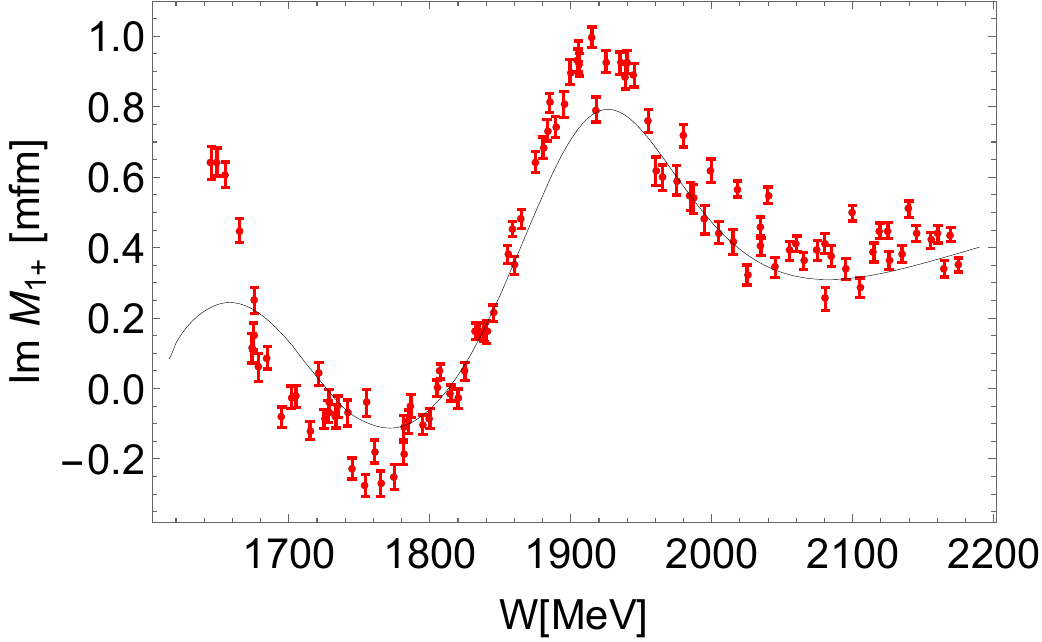}  \\
\includegraphics[width=0.32\textwidth]{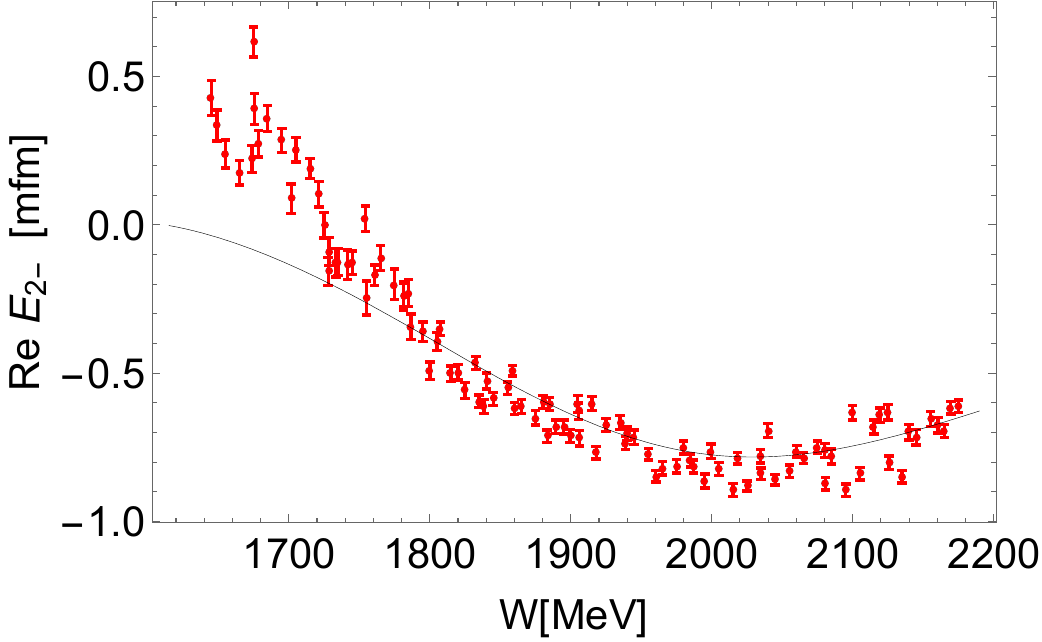} \hspace{0.5cm}
\includegraphics[width=0.32\textwidth]{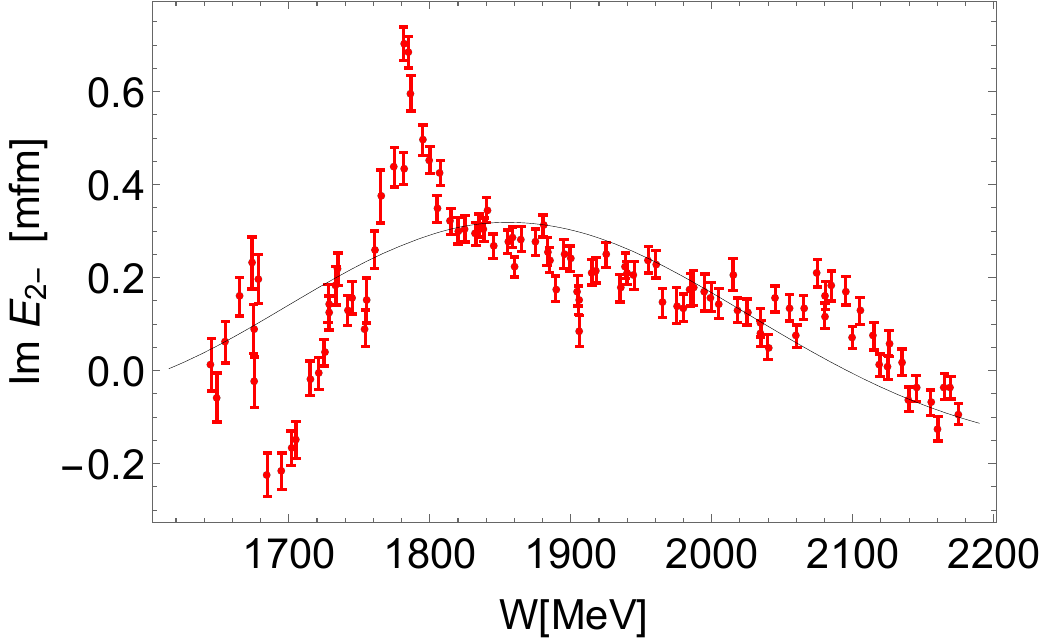}  \\
\caption{\label{MultipolesJB:a}(Color online) The multipoles for the $L=0$, $1$ and $2$ partial waves. Red discrete symbols correspond to our single-channel PWA,  and the full black lines give the JB2017 energy-dependent solution for comparison.  }
\ec
\end{figure}

As before, we also compare the JB2017 energy-dependent total $\chi^2$ values to those obtained in our single-energy fits.
The values of $\chi^2$ per degree of freedom show trends very similar to those we found with BG2017; we again find the
single-energy results oscillating around the expected value of unity.
\begin{figure}[h!]
\bc
\includegraphics[width=0.6\textwidth]{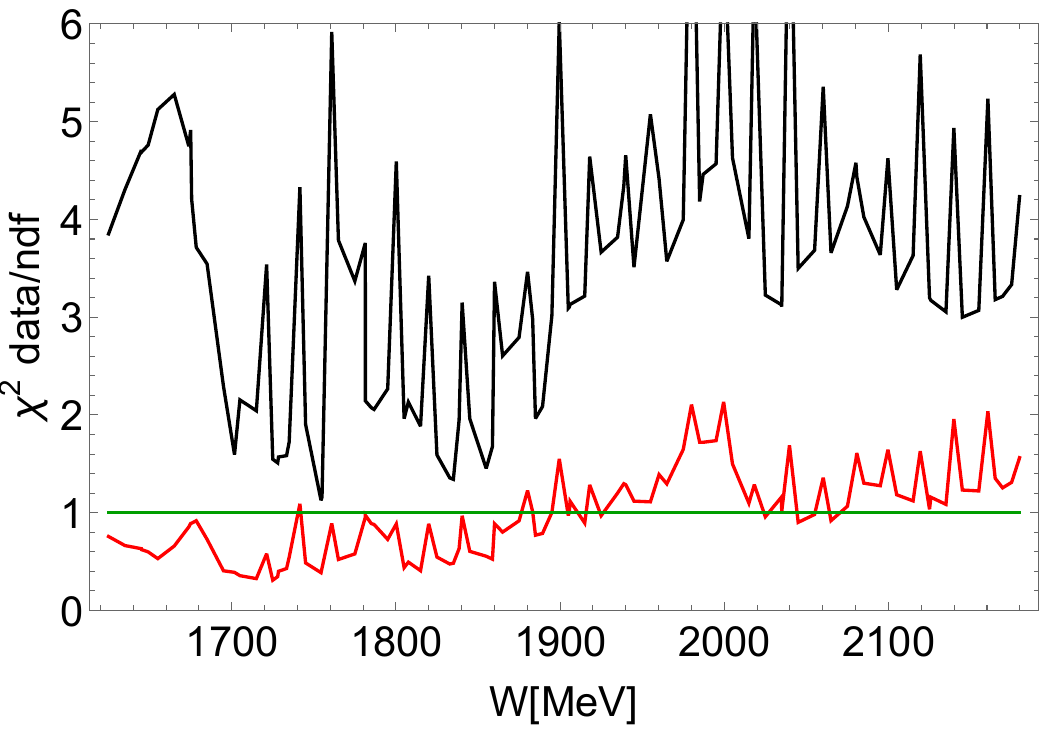}  \\
\caption{\label{chi2JB}(Color online) The obtained $\chi^2$ per degree of freedom.  Black color denotes JB2017 values, and red color denotes our model.}
\ec
\end{figure}
\section{Dependence of final result upon constraining function}
We have tested the influence of the weight of the constraining (penalty) function $\lambda_{ph}$ upon the result, repeating the fits from Sections~\ref{BG2017fit} and \ref{JB2017fit} with increased penalty function weighting factor $\lambda_{ph}$ (see Eq.~\ref{Eq1}). Specifically, we have increased the employed value $\lambda_{ph}=25$ to a much larger value value $\lambda_{ph}=1000$, and established almost invisible differences in the final result. We attributed small differences to the numerical nature of the procedure. So, we conclude that the procedure is independent of $\lambda_{ph}$, and that the used constraining function does not influence observables. And, we believe, this is the closest one can get to theory independent procedure. We have used theory assumptions in choosing the constraining phase, and we used theory assumptions to  generate pseudo-data $E$ and $F$. The latter theory dependence can be eliminated by measuring these pseudo-observables. However, the phase issue can never be eliminated.
\section{Results and Discussion}

Summarizing our results, we have made single-channel, single energy fits to a large database of $K\Lambda$ photoproduction observables (including some pseudo-data). The continuum ambiguity has been resolved using
information from two elaborate, and quite different, multi-channel analyses. We have shown that using the overall phase from either BG2017 or JB2017, in a penalty function, has little to no effect on the single-energy
fit to observables. It does, however, produce multipoles that are in good agreement with the chosen multi-channel fit.
This gives a new way to constrain single-channel single-energy fits and also a simple way to compare the multipoles produced in different energy-dependent fits.

In Figs.~\ref{MultipolesBG:a}~and~\ref{MultipolesJB:a}, for example, we show the  multipoles for $K \Lambda$ photoproduction up to $L_{max}=2$ for constraints from the BG2017 and JB2017 models.
The single-energy points, with errors, are given for each of the interpolated energies. In most cases, they closely follow the smooth BG2017 curves. We observe that both sets of solutions (exactly as their associated theoretical models) are visually very different, but equally well fit the data (see Figs~\ref{chi2BG} and \ref{chi2JB}). So, we can absolutely attribute this difference to the different overall phase.
It should be emphasized that the single-energy multipoles were not constrained to follow BG2017 and JB2017 theoretical amplitudes but rather the constraint was applied to the phase of the $b_1$ transversity amplitude.

\begin{acknowledgments}
We thank D. R\"onchen and L. Tiator for helpful discussions.
A.S. acknowledges the support from STRONG-2020 EU project, Grant agreement ID: 824093.
This work was supported in part by the U.S. Department of Energy, Office of Science, Office of Nuclear Physics,
under grant DE-SC001652.
\end{acknowledgments}
\clearpage
\appendix
\section{} \label{Appendix}

It has been known for decades that $2\to 2$ scattering observables can be represented by complex bilinear expressions and are thus invariant
when each amplitude is multiplied by some energy and angle dependent phase. This is the continuum ambiguity. Thus, if $N$ amplitudes are involved,
they may be expressed in terms of 2N-1 real numbers - the last parameter remains undetermined (as an overall phase).
\\ \\ \noindent
For pseudoscalar meson photoproduction this number is N=4, and all observables are in terms of transversity  amplitudes given in the following table:

\begin{table}[h]
 \begin{center}
 \begin{tabular}{lr}
 \hline
 \hline
  Observable  &  Group  \\
  \hline
  $\sigma_{0} = \frac{1}{2} \left( \left| b_{1} \right|^{2} + \left| b_{2} \right|^{2} + \left| b_{3} \right|^{2} + \left| b_{4} \right|^{2} \right)$  &     \\
  $\hat{\Sigma} = \frac{1}{2} \left( - \left| b_{1} \right|^{2} - \left| b_{2} \right|^{2} + \left| b_{3} \right|^{2} + \left| b_{4} \right|^{2} \right)$  &   $\mathcal{S}$ \\
  $\hat{T} = \frac{1}{2} \left( \left| b_{1} \right|^{2} - \left| b_{2} \right|^{2} - \left| b_{3} \right|^{2} + \left| b_{4} \right|^{2} \right)$  &     \\
  $\hat{P} = \frac{1}{2} \left( - \left| b_{1} \right|^{2} + \left| b_{2} \right|^{2} - \left| b_{3} \right|^{2} + \left| b_{4} \right|^{2} \right)$  &     \\
  \hline
   $\hat{E}  = \mathrm{Re} \left[ - b_{3}^{\ast} b_{1} - b_{4}^{\ast} b_{2} \right]  = - \left| b_{1} \right| \left| b_{3} \right| \cos \phi_{13} - \left| b_{2} \right| \left| b_{4} \right| \cos \phi_{24}$  &  \\
   $\hat{F} = \mathrm{Im} \left[ b_{3}^{\ast} b_{1} - b_{4}^{\ast} b_{2} \right] = \left| b_{1} \right| \left| b_{3} \right| \sin \phi_{13} - \left| b_{2} \right| \left| b_{4} \right| \sin \phi_{24}  $ &  $\mathcal{BT} $ \\
   $\hat{G} = \mathrm{Im} \left[ - b_{3}^{\ast} b_{1} - b_{4}^{\ast} b_{2} \right] = - \left| b_{1} \right| \left| b_{3} \right| \sin \phi_{13} - \left| b_{2} \right| \left| b_{4} \right| \sin \phi_{24} $  &  \\
   $ \hat{H} = \mathrm{Re} \left[ b_{3}^{\ast} b_{1} - b_{4}^{\ast} b_{2} \right] = \left| b_{1} \right| \left| b_{3} \right| \cos \phi_{13} - \left| b_{2} \right| \left| b_{4} \right| \cos \phi_{24} $  &   \\
   \hline
   $\hat{C}_{x'}  = \mathrm{Im} \left[ - b_{4}^{\ast} b_{1} + b_{3}^{\ast} b_{2} \right]  = - \left| b_{1} \right| \left| b_{4} \right| \sin \phi_{14} + \left| b_{2} \right| \left| b_{3} \right| \sin \phi_{23}  $ &  \\
   $\hat{C}_{z'} = \mathrm{Re} \left[ - b_{4}^{\ast} b_{1} - b_{3}^{\ast} b_{2} \right] = - \left| b_{1} \right| \left| b_{4} \right| \cos \phi_{14} - \left| b_{2} \right| \left| b_{3} \right| \cos \phi_{23} $  &  $\mathcal{BR}$   \\
   $\hat{O}_{x'} = \mathrm{Re} \left[ - b_{4}^{\ast} b_{1} + b_{3}^{\ast} b_{2} \right] = - \left| b_{1} \right| \left| b_{4} \right| \cos \phi_{14} + \left| b_{2} \right| \left| b_{3} \right| \cos \phi_{23} $  &  \\
   $\hat{O}_{z'} = \mathrm{Im} \left[ b_{4}^{\ast} b_{1} + b_{3}^{\ast} b_{2} \right] = \left| b_{1} \right| \left| b_{4} \right| \sin \phi_{14} + \left| b_{2} \right| \left| b_{3} \right| \sin \phi_{23} $  &   \\
   \hline
   $\hat{L}_{x'} = \mathrm{Im} \left[ - b_{2}^{\ast} b_{1} - b_{4}^{\ast} b_{3} \right] = - \left| b_{1} \right| \left| b_{2} \right| \sin \phi_{12} - \left| b_{3} \right| \left| b_{4} \right| \sin \phi_{34}$  &   \\
   $\hat{L}_{z'}  = \mathrm{Re} \left[ - b_{2}^{\ast} b_{1} - b_{4}^{\ast} b_{3} \right]  = - \left| b_{1} \right| \left| b_{2} \right| \cos \phi_{12} - \left| b_{3} \right| \left| b_{4} \right| \cos \phi_{34}$  &  $\mathcal{TR}$  \\
   $\hat{T}_{x'} = \mathrm{Re} \left[ b_{2}^{\ast} b_{1} - b_{4}^{\ast} b_{3} \right] = \left| b_{1} \right| \left| b_{2} \right| \cos \phi_{12} - \left| b_{3} \right| \left| b_{4} \right| \cos \phi_{34}$  &    \\
   $\hat{T}_{z'} = \mathrm{Im} \left[ - b_{2}^{\ast} b_{1} + b_{4}^{\ast} b_{3} \right] = - \left| b_{1} \right| \left| b_{2} \right| \sin \phi_{12} + \left| b_{3} \right| \left| b_{4} \right| \sin \phi_{34}$ &   \\
   \hline
   \hline
 \end{tabular}
 \end{center}
  \caption{The definitions of the $16$ polarization ob\-serva\-bles of pseudoscalar meson photoproduction
 are given here in terms of transversity amplitudes $b_{1}, \ldots, b_{4}$ (
 sign conventions are consistent with~\cite{Bonn}). Expressions are also given in terms of  moduli and relative phases of the amplitudes.
 Furthermore, the phase-space factor $\rho$ has been suppressed in the given expressions (i.e. we have set $\rho = 1$).
 The four different groups of four observables each are indicated as well.}
 \label{tab:PhotoproductionObservables}
\end{table}
\newpage
With this definition we see that single-channel observables $d\sigma/d\Omega$, $\Sigma$, T and P  actually exactly give four absolute values, and from the rest of the spin observables
another three relative angles are determined. Our experiments give us absolute values and sines and cosines of all relative angles, so we need more than seven observables to be measured. We see that by measuring 12 observables we can certainly determine all seven quantities (4 absolute values and 3 relative angles) exactly. Applying discrete symmetries it has been shown that only 8 observables suffice. However, the overall phase can never be determined. In the present study, we avoid the complete-experiment issue by always considering cases where more than a sufficient number of observable types is used.
\\ \\ \noindent
Now we can express transversity amplitudes in terms of 4 absolute and 3 relative angels (measurable quantities), and separate out the overall phase which cannot be measured. This can be done in at least four different ways:
\begin{table}[h]
\begin{tabular}{lclclclcl}
$b_1 =  |b_1|  \cdot {\red e^{i \Phi_{1}}}$ & & $b_1 =   |b_1| \cdot e^{i \Phi_{12}} \cdot  {\red e^{i \Phi_{2}}}$ & & $b_1 =  |b_1| \cdot e^{i \Phi_{13}} \cdot {\red e^{i \Phi_{3}}}$ & & $b_1 =  |b_1| \cdot e^{i \Phi_{14}} \cdot {\red e^{i \Phi_{4}}}$  \\
$b_2 =  |b_2| \cdot e^{i \Phi_{21}} \cdot {\red e^{i \Phi_{1}}}$ & \hspace*{0.5cm} or \hspace*{0.5cm} & $b_2 =  |b_2| \cdot  {\red e^{i \Phi_{2}}}$  & \hspace*{0.5cm} or \hspace*{0.5cm} & $b_2 =  |b_2| \cdot e^{i \Phi_{23}} \cdot {\red e^{i \Phi_{3}}}$  &  \hspace*{0.5cm} or \hspace*{0.5cm} & $ b_2 =  |b_2| \cdot e^{i \Phi_{24}} \cdot {\red e^{i \Phi_{4}}}$  \\
$b_3  = |b_3| \cdot e^{i \Phi_{31}} \cdot {\red e^{i \Phi_{1}}}$ & &$b_3 =  |b_3| \cdot e^{i \Phi_{32}} \cdot {\red e^{i \Phi_{2}}}$  & & $b_3 =  |b_3|  \cdot {\red e^{i \Phi_{3}}}$  & & $b_3 =  |b_3| \cdot e^{i \Phi_{34}} \cdot {\red e^{i \Phi_{4}}}$   \\
$b_4  = |b_4| \cdot e^{i \Phi_{41}} \cdot {\red e^{i \Phi_{1}}}$ & &$b_4 =  |b_4| \cdot e^{i \Phi_{42}} \cdot {\red e^{i \Phi_{2}}}$  & & $b_4 =  |b_4| \cdot e^{i \Phi_{43}} \cdot {\red e^{i \Phi_{3}}}$  & & $b_4 =  |b_4| \cdot {\red e^{i \Phi_{4}}}$   \\
\end{tabular}
\caption{Transversity amplitudes in terms of absolute values and relative angles when overall phase is chosen to be one of the transversity amplitude angles.}
\label{tab:Relamp1}
\end{table}

In each column a different overall phase is separated ($\Phi_1$, $\Phi_2$, $\Phi_3$, or $\Phi_4$) from the rest of the formula which can be exactly extracted from single-channel measurements. In an ideal case, all four ways are equivalent, and we can chose the most convenient one. This set of formulas tells us exactly that reaction amplitudes are undetermined without fixing the overall phase of either of the 4 amplitudes to some chosen, predetermined value.

\end{document}